# Mokka: BFT consensus

**Zuev Egor**

Higher School of Economics

zyev.egor@gmail.com

## Abstract

Mokka is a partial-synchronous, strong consistent BFT consensus algorithm for reaching the consensus about a certain value in open networks. This algorithm has some common approaches nested from RAFT, but its nature and design make Mokka a better solution for DLT (distributed ledger).

## 1. Introduction

Mokka is a partial-synchronous, strong consistent BFT algorithm. The algorithm implements consensus by first electing a single leader, then giving the leader full responsibility for the global state (i.e., state mutation). This approach is also known as leader-follower model. Mokka is a log-less algorithm, which means there is no log replication out-of-the-box, but this behavior is extendable (section 3). The follower acts passively: they listen to the leader and accept global state changes (if implemented log). If the leader goes down, then the new leader is elected. The BFT properties of Mokka are based on digital signatures (elliptic curve SECP256K1) and offer both liveness and safety for at most $N = 3f + 1$ with quorum of $Q = 2f + 1$, where N – the amount of nodes in cluster and $f$ – is the number of faulty nodes.

As a result, following the leader-follower model and BFT properties, Mokka aims to solve the consensus problem, where a Byzantine fault is possible.

#### 2.2 System model

The system represents a partially synchronous distributed system, where nodes are connected by the network. We assume that the network may duplicate, fail to deliver, or deliver with delays messages. We also assume that network can be split into two or more networks. Each node in the network may become failed. By failed node, we mean arbitrary behavior or out of order (when a node goes offline). To maintain system operatable, there are required $N = 3f + 1$ nodes, where f – is the amount of possible failed nodes.

The system is partially synchronous as the timers are used to restrict leader election, detect leader down and limit the leadership of the certain node (to prevent network freeze).

The algorithm uses the leader-follower model, which means there is no intercommunication between followers (like happen in PBFT). The only communication happens from leader to follower (to maintain the leader state) and between candidate and follower node (to make voting happen). The algorithm relies on the voting process to elect a new leader and uses multi-signatures to prove that voting was done only by authorized and known nodes in the system.

To maintain its state, the leader periodically sends ACK packets to followers. If the packet has not been sent by the leader or received by a follower in a certain time frame (called heartbeat timeout), followers assume that the leader was down and a new leader should be elected.

The leader uses an aggregated token (built using multi-signature, term, and nonce) as an authentication and authorization token when sending ACK packets to followers. The nonce – represents the timestamp when voting started. Using nonce and setting proof expiration, we control the leadership limit (to prevent a single malicious leader from holding the network). Also, the algorithm requires changing the leader between $2f + 1$ nodes in a row to prevent faulty nodes from passing the leadership between them and allowing at least one healthy node to take place as leader.

Each voting round starts with a new term. The term should be incremented per each new voting round. The system guarantee that there will always be one leader per term. To prevent the system from early voting (to prevent federation of malicious nodes vote for malicious leader skipping healthy node), the algorithm uses “voting lock”, which locks voting until next ACK from the current leader or heartbeat timeout (to make sure that current leader was down).

The algorithm guarantees **safety and liveness** as long as $N \geq 3f + 1$, and the global clock is synchronized. The safety property means that all healthy nodes agree on the same term and leader. The liveness means that even if $f$ nodes fail, the network will still be operatable, and the cluster will elect a new leader.

### 2.3 algorithm basics

Mokka is a partial-synchronous, strong consistent BFT algorithm which use Leader-follower model to reach consensus. There are three roles presented: leader, follower, the candidate. At any time point, the node should be in one of these states. The leader is chosen through the election process (i.e., voting), and in normal operation – there should be one leader.

Each voting happens around the term. The term represents a logical clock. Each time new voting happens, the term should be incremented. In this way, over each term, there can be at most one leader. Under certain conditions, like network split, there possible to have two leaders, but under different terms. Mokka use timeouts for leader, so even if the situation happens with two leaders over different terms in a split network, the leader won’t exist forever and have to transit to follower state, once leader timeout happens. Also, it is possible not to have a leader over a certain term. This happens when a node doesn’t receive enough votes to reach the quorum. The quorum is 51% of all nodes, including those who want to become the leader.

Each node in the network should have its own private + public keys pair, generated by SECP256K1 specification. Mokka uses private and public keys during the voting process – the vote should be signed by the corresponding node. Then these votes are used as part of multi-signature and as proof of voting for the certain leader in the network over the specific term. Also, the public + private key pair can be used to sign the packets (section 3), but this behavior is optional.

The algorithm communication is represented as RPC and includes three types of messages (packets): ACK, VOTE, and VOTED.

### 2.4 Initial state

Before the boot phase, each node should have the following properties (or settings):

| Property | Description | Example |
|---|---|---|
| Private key | generated with secp256k1 standard | “edca08af903bbd7afd356bb2d6fdf54901fad8770c734416456d382012b0608b” |
| Public key | generated with secp256k1 standard (from private key) | “02b68e80e752f13906179688171d6a0345a6c6edd20b52f24ea489275668fa1eac” |
| Election timeout | Election timeout (ms) | 100 |
| Heartbeat | Heartbeat timeout (ms) | 50 |
| Proof expiration | How long the proof is valid (ms) | 300 |
| Nodes | An array of nodes (peers) addresses (should include network address + public key of the peer) | [<br>“tcp://127.0.0.1:2000/036032cc7790000ac98fc2dbc200c486b073e74e8e4ff5fba15beb01c153f4c458”<br>] |

Figure 2. Settings for the node

### 2.5 Boot phase

During the boot phase (when we have a bootstrapped list of all nodes in the network) or when the node adds new nodes to the list of known nodes, the node generates/regenerates the $rootPublicKey$ and combinations of public keys. The $rootPublicKey$ is a summarization of all known public keys in the network. The addition happens with point math:

$$rootPublicKey = \sum publicKey_i$$

Formula 1. Root public key

Where i – is an index of a node.

For instance, if the network has nodes A, B, and C, then:

$$rootPublicKey = publicKey_A + publicKey_B + publicKey_C$$

The $rootPublicKey$ should be the same across all nodes, and this will help ensure that all nodes in the network have the same set of bootstrapped nodes, and as a result – the same set of public keys.

The next step is to generate the combinations of public keys. To do that, the node should find all combinations of public keys by a quorum in sorted order (to exclude the same pairs with different order). For instance, if the network has nodes A, B, and C, then pairs will be: (A, B), (B, C), (A, C).

**2.6 Leader election**

Mokka uses a heartbeat mechanism to trigger a leader election. When nodes startup, they begin as followers. A node remains in a follower state as long as it receives a valid ACK packet from the leader. Leader multicasts ACK packets to all followers periodically. This interval is called heartbeat timeout. The follower listens to ACK packets and has its own timeout. If there is no communication within the timeout (it's called safe timeout), then the follower assumes there is no viable leader and begins an election to choose a new leader. If the follower receives ACK within a defined timeout, then the timer drops and starts over. The follower's timeout should be bigger than the leader's heartbeat (as network lags are possible) and be random (to overcome the problem when all followers simultaneously get timeout). Mokka uses the following formula to obtain follower's timeout:

$$timeout = heartbeat\ timeout * 1.5 + random(1, heartbeat\ timeout)$$

Formula 2. follower timeout

To begin an election, a follower increments its current term and transitions to a candidate state. Then the candidate should generate the nonce, $rootPublicKey\mathrm{ForTerm}$, $sharedPublicKeys$.

The nonce – is a current timestamp. The nonce should be unique per each voting. There are two main reasons for introducing nonce: to make hashes for signatures unique (to avoid replay attack) and to introduce the leader timeout (to avoid network freeze). The nonce is obtained as a timestamp:

$$nonce = current\ timestamp$$

Formula 3. Nonce

The next step is to generate $rootPublicKey\mathrm{ForTerm}$ – this is a unique public key, which will be used as part of shared public keys. The idea of $rootPublicKey\mathrm{ForTerm}$ is to introduce the unique public key, which is a combination of the nonce, current term, and candidate's public key. Such combination will help followers ensure that a certain shared public key has been generated for the certain node over a certain term and nonce. This will eliminate replay attacks and fake leaders:

$$rootPublicKey\mathrm{ForTerm} = rootPublicKey * HASH(nonce\ ||term\ ||publicKey)$$

Formula 4. Root public key for the term

Where $publicKey$ is a public key of the candidate, multiplication operation – is a point multiplication, and HASH function is SHA256.

Then the candidate will have to generate shared public keys for each of the public keys combinations where combination size in equal to quorum size and where his public key is present, generated on boot step:

$$sharedPublicKey_X = \sum publicKey_i * HASH(nonce \mid\mid term \mid\mid rootPublicKey\text{ForTerm})$$

Formula 5. Shared public key

Where x – is a combination of public keys, $publicKey_i$ – is a public key of i node, and HASH function is SHA256. The $sharedPublicKey$ should be generated for each combination of public keys.

For instance, if the network has nodes A, B, C and D, then there will be three pairs for node A – (A, B, C), (A, C, D), (A, B, D); three pairs for node B – (A, B, C), (B, C, D), (A, B, D); and three pairs for node C – (A, B, C), (B, C, D), (A, C, D). From the view of node A, the pairs will be generated like:

$$sharedPublicKey_{A,B,C} = (publicKey_A + publicKey_B + publicKey_C) * HASH(nonce \mid\mid term \mid\mid rootPublicKey\text{ForTerm})$$

$$sharedPublicKey_{A,C,D} = (publicKey_A + publicKey_C + publicKey_D) * HASH(nonce \mid\mid term \mid\mid rootPublicKey\text{ForTerm})$$

$$sharedPublicKey_{A,B,D} = (publicKey_A + publicKey_B + publicKey_D) * HASH(nonce \mid\mid term \mid\mid rootPublicKey\text{ForTerm})$$

The second step is to vote for itself (as a candidate). To do that, we have to generate the partial signature:

$$partialSignature = privateKey * HASH(nonce \mid\mid term \mid\mid rootPublicKey\text{ForTerm}) \bmod n$$

Formula 6. Partial signature

Where n – is a curve (SECP256K1) parameter. Modulo n is an optional step, which will reduce the signature size.

The last step is to multicast the VOTE packet to each node. Here is the structure of the VOTE packet:

| Field | Description | Example value |
|---|---|---|
| Type | The packet type | "VOTE" |
| State | The node's state. Can be candidate, leader, or follower | "candidate" |
| Term | The node's term, who send a packet | 1 |
| publicKey | The node's public key, which sends a packet | "036032cc7790000ac98fc2dbc200c486b073e74e8e4ff5fba15beb01c153f4c458" |
| Nonce | The candidate's generated nonce | 1628061165517 |

Figure 3. VOTE packet structure

Now each follower, who receives the VOTE packet should vote for the following candidate. The follower performs several checks.

During the first check, the algorithm validates that the candidate's term is higher than the current one, and it validates the nonce value that it is not expired:

$$current\ timestamp - nonce < \text{Proof expiration}$$

Figure 4. nonce expiration

Then the follower should validate the term (that it's not too high). To make this happen, the algorithm should store the timestamp of the last state update (when the term was incremented). By knowing that leader timeout minimum time is $heartbeat\ timeout * 1.5 + 1$, and the minimum election time is higher than the heartbeat, then even if leader switch will happen in 1ms, the timeout and election of a new leader will start not earlier than $heartbeat\ timeout * 1.5 + 1$ of time. From this point, the highest term can be calculated as:

$$highest\ term = current\ term + (current\ timestamp - last\ term\ update\ timestamp)/heartbeat$$

Formula 7. Highest possible term

The next validation checks that there is a new leader in this term. As the algorithm should be resistant to arbitrary behavior from f nodes, each follower should store the last f term's leaders. For instance, if we have f = 2, the $N = 3f + 1 = 3 * 2 + 1 = 7$ consisting of nodes (A, B, C, D, E, F, G), the current term = 3, the term = 1 leader was node A, the term = 2 leader was node B, then for term = 3 the leader node can be C, D, E, F or G. In the worst scenario, if A and B nodes are arbitrary nodes, the algorithm guarantees that one node will still serve requests, but as arbitrary nodes may not accept the client's requests, the network speed may degrade.

The last validation is to check that the current leader is down. To prevent the follower from early voting (when the current leader is still present, but an arbitrary node asks for leader change), the follower makes a "voting lock". Voting lock – locks the voting function until ACK or timeout happens. The timeout should be bigger than the leader's heartbeat timeout (as network lags are possible). The optimal value for such timeout is $heartbeat\ timeout * 1.5$. If the ACK has been received and the token is still valid, the follower doesn't vote for leader change. Otherwise, the algorithm process further.

If any of the checks failed, the follower sends back the VOTED packet, providing the null value for signature. Otherwise, the follower generates publicKeysRootForTerm and partial signature, then updates its term to the candidate's version.

The publicKeysRootForTerm (Formula 4. Root public key for the term) is generated using the candidate's public key, nonce, and term. All of them are retrieved from the VOTE packet. The generated publicKeysRootForTerm by the follower should be equal to the candidate's one.

Partial signature (Formula 6. Partial signature) is generated using candidate's nonce and term, generated publicKeysRootForTerm and its (follower) own private key.

The last step is to send a reply VOTED packet to the candidate. Here is the structure of voted packet:

| Field | Description | Example value |
| --- | --- | --- |
| Type | The packet type | "VOTED" |
| State | The node's state. Can be candidate, leader, or follower | "follower" |
| Term | The node's term, who send a packet | 1 |
| publicKey | The node's public key, which sends a packet | "02b68e80e752f13906179688171d6a0345a6c6edd20b52f24ea489275668fa1eac" |
| Signature | Partial signature | "3cb982501b4deec297cf71b6d8fe85aed19a9aa0b4b68c58e070a9ecd9b58945" |

Figure 5. VOTED packet structure

In case the candidate receives enough votes (quorum), he builds a full signature, which is also a multi-signature:

$$signature = \sum partialSignature_i$$

Formula 8. signature

The thing about the full signature is that it consists of N partial signatures, where N – is a quorum. For instance, in a network with nodes A, B, C and D where A is a candidate, the full signature will consist of either (A, B, C) or (A, C, D) or (A, B, D) partial signatures. Let's assume nodes B and C voted for node A; then the full signature will look like that:

$$signature_{A,B,C} = partialSignature_A + partialSignature_B + partialSignature_C$$

Next, by knowing which nodes participated in the voting, we can find the corresponding shared public key generated earlier (Formula 5. Shared public key).

The last step is to build a special proof, which will act as auth token when the leader will send ACK packets. To do that, concatenate the nonce, shared public key, and signature:

$$proof = nonce||sharedPublicKey||signature$$

Formula 9. Proof token

**To summarize the key concept:**

1) Each node generates the root public key from all known public keys. The result should be the same on all nodes. This helps ensure that all nodes know their neighbors (and there is no fake/unknown node) and have the same quorum requirements (for instance, 5 of 7 should vote).
2) Each node (candidate and followers) builds the coefficient, which is unique per nonce, term, and public key. The generated coefficient should be the same on all nodes. This helps ensure that all nodes vote for a certain candidate over a certain term in the certain network (as coefficient use root public key).
3) The candidate generates shared public keys, including only pairs, where the candidate's public key is present. It is a multi-public key of (M-of-N) nodes. For instance, in a cluster of A, B, C and D nodes, where node A is a candidate, the possible multi-public keys are (A, B, C), (A, C, D), (A, B, D). Technically, when voting happens, to become the leader, we need to obtain a vote (partial signatures) from either (A, B, C) or (A, C, D) or (A, B, D) combination.
4) During the voting process, followers sign a generated hash of root public key for term, term, nonce, and candidate's public key. This is called a partial signature. When the candidate received enough votes (partial signatures), he combines them (addition operation).
5) It is fair to say that the shared public key is equal to signature * G (the proof is presented in the next section)

That was a successful flow. Other outcomes are also possible: another node becomes a leader, election timeout triggers, and leader fallback to follower state.

The first negative scenario is related to separated voting when several candidates are possible at the same timeframe and even over different terms. Mokka uses randomized election timeouts within a predefined time-box (1.5x-2.5x of heartbeat timeout) to prevent multiple candidates in the same voting round. Even if one of the candidates will become the leader, another candidate will fall back to a follower state.

The second negative outcome is related to election timeout. This is possible when there are too many candidates or thanks to network lags. For this purpose, it is recommended to log such events to ensure there is no network degradation; otherwise, the consensus won't be possible.

### 2.7 Leader validation

After the candidate becomes a leader, he starts sending the ACK packets. The structure of the ACK packet is the following:

| Field | Description | Example value |
|---|---|---|
| Type | The packet type | "ACK" |
| State | The node's state. Can be candidate, leader, or follower | "leader" |
| Term | The node's term, who send a packet | 1 |
| publicKey | The node's public key, which sends a packet | "02b68e80e752f13906179688171d6a0345a6c6edd20b52f24ea489275668fa1eac" |
| Proof | The proof of voting | "1628067685604:02d9e4035dccba9932bf37de6a35b60cda78e8531d2b2c1df602d1a534483ae0d4:1b4dd986cd69edee45078481fb99f0ce102de3202b2b127ac2a107d0b0a9f8a06" |

Figure 6. ACK packet structure

On first receiving an ACK from the leader for the current term, each follower should validate the proof to ensure that the current leader is a real one. The validation happens in 3 levels:

1) validation of nonce against proof expiration (Figure 4. nonce expiration)
2) validation of shared public key (that the follower recognizes this shared public key)
3) validation of signature against the shared public key

By knowing the proof and its structure, a follower can extract the nonce, shared public key, and signature. To validate the shared public key, the follower should:

1) generate $rootPublicKey\text{ForTerm}$ (Formula 4. Root public key for the term), where nonce, term, and publicKey are obtained from ACK packet
2) build a shared public key for all known combinations of public keys (Formula 5. Shared public key)
3) check if supplied shared public key exist among generated. If there is no such – then the leader is not authorized, and his packet won't be processed

The last validation is a validation of signature against shared public key:

$$signature * G = sharedPublicKey$$

Formula 10. Signature validation

Where G – is a curve parameter (point).

In case of equality is passed, then we assume that the leader is authorized. Here is the proof:

$$signature * G = sharedPublicKey$$

$$\sum partialSignature_i * G = \sum publicKey_i * HASH(nonce \mid\mid term \mid\mid rootPublicKey\text{ForTerm})$$

$$\sum (privateKey_i * \ HASH(nonce \mid\mid term \mid\mid rootPublicKeyForTerm)) \bmod n * G$$
$$= \sum publicKey_i * HASH(nonce \mid\mid term \mid\mid rootPublicKey\text{ForTerm})$$

$$\sum privateKey_i * G = \sum publicKey_i$$

Formula 11. Proof validation

If the token is wrong or outdated, the follower will stop accepting ACK packets from the leader and propose itself as a candidate.

### 2.8 Safety

The safety property guarantees, that under different circumstances, the state will be maintained by the quorum and correctly handled by state machine.

The algorithm guarantees that:

1) there will be only one leader per term
2) The single leader can't remain forever due to leadership expiration
3) The $f$ nodes will not take the control over the cluster by passing the leadership to each other, as rotation algorithm requires that at least $Q - 1$ nodes will take the leadership before the the fail node can become the leader again.
4) Only authenticated nodes can vote for the next leader. The result of voting – is a multisig with known indexes of participants, which makes it possible to detect, which nodes exactly voted, and make it possible for other followers to validate multisig and current leader
5) The safety guarantees by at least $N = 3f + 1$

As current algorithm doesn't propose any RSM implementation, nor logs, we won't cover the corresponding safety properties for logs replication and state machine execution order.

#### 2.8.1 Safety extra assumptions (2f + 1)

Since Mokka propose only reaching the quorum and select the leader through voting and roation mechanizm, and doesn't propose any state management, it's fair to say, that BFT properties can be reached by using $N = 2f + 1$, or quorum of $f + 1$. In this case, the worst scenario, which may happen, is that two or more nodes may be the leader at the same term, but at different time point.

# 2. Extensions to the original algorithm

### 3.1 About

In this section, some ideas are discussed regarding extending Mokka's functionality.

### 3.2 Logs

Mokka can be extended and support logs. This can be achieved in 2 ways: by extending the Mokka or implementing the logs on level 2.

The first approach is related to writing an extension to Mokka. As Mokka has a lot in common with RAFT, extending Mokka according to RAFT's whitepaper is recommended. This will add logs support and some rules to voting (like validating the last log index).

Another approach is about writing the logs support on application level (or level 2). In this way, synchronizing changes between leader and follower will be isolated from the consensus algorithm.

### 3.3 Packet digital signatures

To maintain the leader's authority and have a chance to detect the leader's replacement immediately, packet's digital signatures can be introduced. The concept is in signing each sent packet and validate it on the receiver side.

The signature should be created by the hash of the original packet (which node is going to send). To make packets unique (to avoid replay attack), it is recommended to introduce the timestamp and compare it against proof expiration time (or you can specify a different timeout value). Also, even if a packet has a timeout, remember that some packets, like VOTE – have the same body so that they can be easily reused within the specified time box. To avoid reuse, the algorithm can be extended further and introduce the packet index. On each new term, the algorithm will store the index of each sent/received packet per node. In this case, every new packet should have a higher index than the previous one. This rule should be checked before processing the packet. This flow will make the packet unique per time (time box), term, and index.

The downside of this approach is that the signing and validation process of digital signature requires some time (these operations are not lightweight), which means that there will be delays in packet processing, which will lead to network degradation or network hold (as ACK packets will be delivered and processed with delays and timeout on follower may trigger at this time). For this reason, we've excluded this functionality in Mokka by default.

# Conclusion

In the following article, we've proposed the consensus BFT algorithm Mokka; its architecture; and discovered its properties. Mokka introduced a novel way to voting and authorizing the leader by using a multisig approach. This approach is scalable as proof size is static (no matter how many nodes are present in the network). Also, the algorithm's properties make it easily extendable and lightweight comparing to other classical BFT algorithms.

The implementation of Mokka consensus is available in our GitHub repository[1].

## References

1. Mokka GitHub repository: https://github.com/ega-forever/mokka
2. Cynthia Dwork, Nancy Lynch and Larry Stockmeyer:
   Consensus in the Presence of Partial Synchrony -
   https://groups.csail.mit.edu/tds/papers/Lynch/jacm88.pdf
3. Denis Rystsov. CASPaxos: Replicated State Machines without logs -
   https://arxiv.org/pdf/1802.07000.pdf
4. Diego Ongaro and John Ousterhout. In Search of an Understandable Consensus Algorithm -
   https://raft.github.io/raft.pdf
5. Robbert van Renesse, Dan Dumitriu, Valient Gough, Chris Thomas. Efficient Reconciliation and -
   Flow Control for Anti-Entropy Protocols -
   http://www.cs.cornell.edu/home/rvr/papers/flowgossip.pdf
6. Daniel R. L. Brown. Recommended Elliptic Curve Domain Parameters -
   http://www.secg.org/sec2-v2.pdf
7. Colin J. Fidge. Timestamps in Message-Passing Systems That Preserve the Partial Ordering-
   http://fileadmin.cs.lth.se/cs/Personal/Amr_Ergawy/dist-algos-papers/4.pdf
8. A. Shamir. How to share a secret”, Communications of the ACM 22 (11): 612613, 1979.
9. The Elliptic Curve Diffie-Hellman (ECDH) -
   https://koclab.cs.ucsb.edu/teaching/ecc/project/2015Projects/Haakegaard+Lang.pdf
10. Distributed systems for fun and profit - http://book.mixu.net/distsys/single-page.html
11. Practical Byzantine Fault Tolerance and Proactive Recovery -
    http://www.pmg.csail.mit.edu/papers/bft-tocs.pdf
12. CONSENSUS: BRIDGING THEORY AND PRACTICE - https://web.stanford.edu/~ouster/cgi-bin/papers/OngaroPhD.pdf